\documentclass[epj]{svjour}
\usepackage{graphics}
\usepackage{graphicx}
\usepackage{bbm}
\usepackage{amsmath,amssymb}

\newcommand{\be}{\begin{equation}}
\newcommand{\ee}{\end{equation}}
\newcommand{\bea}{\begin{eqnarray}}
\newcommand{\eea}{\end{eqnarray}}

\begin{document}
\title{Optimal control of quantum revival}
\author{Esa R\"as\"anen\inst{1,2,3} \and Eric J. Heller\inst{3,4}
}                     
%
%
\institute{Department of Physics, Tampere University of Technology, FI-33101 Tampere, Finland \and Nanoscience Center, Department of Physics, University of Jyv{\"a}skyl{\"a}, FI-40014 Jyv{\"a}skyl{\"a}, Finland \and Physics Department, Harvard University, Cambridge, Massachusetts 02138, USA \and Department of Chemistry and Chemical Biology, Harvard University, Cambridge, Massachusetts 02138, USA}
\date{Received: date / Revised version: date}
%
\abstract{
Increasing fidelity is the ultimate challenge of
quantum information technology. In addition to decoherence
and dissipation, fidelity is affected by internal 
imperfections such as impurities in the system. 
Here we show that the quality of quantum revival,
i.e., periodic recurrence in the time evolution,
can be restored almost completely by coupling the 
distorted system to an external field obtained from quantum 
optimal control theory. We demonstrate the procedure 
with wave-packet calculations in both one- and 
two-dimensional quantum wells, and 
analyze the required physical characteristics of the 
control field. Our results generally show that the inherent
dynamics of a quantum system can be idealized 
at an extremely low cost.
} 
\maketitle
\section{Introduction}

Quantum revival, i.e., periodic recurrence of a wave function 
is a fundamental property of a time-dependent quantum 
system.  The quality of the quantum revival can be determined
by a simple overlap between the initial and final 
states after the {\em ideal} revival time, i.e.,
the revival time of the unperturbed system.
In this respect, the quality corresponds to quantum 
fidelity~\cite{review}.

Fidelity is the most important measure of the 
functionality of a quantum information device,
and in this respect it has been subjected to
extensive theoretical work~\cite{kohler}.
In experiments fidelity is greatly affected
not only by eventual decoherence and dissipation due 
to the coupling with the environment, but
also by internal imperfections such as random
irregularities and impurities. The standard approach
to overcome this problem is to increase the sample 
quality or to make it, by other means, particularly 
resistant to imperfections. 

An alternative strategy to improve fidelity, which is
the topic of this work, is to couple the system to an external
{\em control field} that assists the system to 
overcome and/or to compensate the effects induced by
irregularities. Such scenarios have attracted significant
interest in the design of high-fidelity quantum gates
in optical lattices~\cite{khani}.
Here we show that optimizing the control
field with quantum optimal control theory~\cite{oct1,oct2} (OCT) 
leads to a practically complete restoration of the fidelity
and thus to the full quantum revival even at 
significantly low energies of the control field. 
We analyze in detail the properties of the optimized 
field and demonstrate the scheme for the time evolution 
of one- and two-dimensional wave packets, respectively.
Our findings have broader applicability to general control
problems, where the inherent dynamics can be assisted by optimized 
fields subject to strict constraints in terms 
of the intensity and the frequency range.

\section{Optimal control theory}

Since its formulation~\cite{oct1,oct2} in the 1980s OCT has increased its
popularity in chemistry and condensed matter physics~\cite{rabitz}.
The central idea of OCT is to replace trial-and-error type 
learning-loop experiments with a rigorous extension 
of the classical control problem to quantum mechanics.
In all OCT applications the objective is to find 
an external time-dependent field 
$\boldsymbol{\epsilon}(t)$ that drives the 
system to the predefined target, e.g., to a certain
quantum state, through
the solution of the Schr\"odinger equation,
\begin{equation}\label{schr}
i\frac{\partial}{\partial t}\Psi(\boldsymbol{r},t)=
\left[\hat{T}+\hat{V}-\hat{\mu}\boldsymbol{\epsilon}(t)\right]\Psi(\boldsymbol{r},t),
\end{equation}
where dipole approximation is applied with $\hat{\mu}=-{\mathbf r}$
(Hartree atomic units used throughout).
The central idea is to maximize the target functional
\begin{equation}
\label{j1}
 J_{1}[\psi]=\left<\Psi(\boldsymbol{r},T)|\hat{O}|\Psi(\boldsymbol{r},T)\right>,
\end{equation}
at time $t=T$. In this study we consider two types of target operators: 
(i) projection operator 
$\hat{O}=\left|\Phi_{\rm F}\right>\left<\Phi_{\rm F}\right|$ 
with $\Phi_{\rm F}$ as our target {\em state}, and (ii) a local operator
$\hat{O}=\rho_{\rm F}({\mathbf r})$ representing our target 
{\em density} at $t=T$. 
The maximum value obtained for $J_1$ is referred to the {\em yield}.

The optimized pulse is exposed to two important constraints.
First, the fluence, i.e., the time-integrated intensity $F_0$
is kept fixed by applying a functional,
\begin{equation}
 J_{2}[\epsilon]=-\alpha\left[\int_{0}^{T}dt\,\epsilon^{2}(t)-F_0\right],
\end{equation}
where $\alpha$ is a time-independent Lagrange multiplier.
Secondly, a spectral filter is applied to cut off the highest
frequencies, e.g., unrealistic photon energies.
This is done by multiplying the Fourier-transformed pulse
by a filter function $f(\omega)$, that represents the
desired frequency range. Thereafter, an inverse Fourier transform
of the product is taken to obtain the time-resolved 
filtered pulse (see Ref.~\cite{janreview} for details).

Finally, the time-dependent Schr\"odinger equation [Eq.~(\ref{schr})] 
must be satisfied in the control procedure. This gives us yet another 
functional,
\begin{equation}
  J_{3}[\epsilon,\Psi,\chi] = -2\,\textrm{Im}\int_{0}^{T}\big<\chi(t)|i\partial_{t}-\hat{H}(t)|\Psi(t)\big>,
\end{equation}
where the auxiliary function 
$\chi(t)$ can be regarded as a time-dependent Lagrange 
multiplier~\cite{oct1}.

Variation of the total functional $J=J_{1}+ J_{2}+J_{3}$ with respect 
to $\Psi$, $\chi$, $\epsilon$, and $\alpha$ leads 
to the control equations
\begin{eqnarray}
	i\partial_{t} \Psi(t) & = & \hat{H}(t)\Psi(t), \quad \Psi(0)=\Phi_{I},\\
	i\partial_{t} \chi(t) & = & \hat{H}(t)\chi(t), \quad \chi(T)=\hat{O}\Psi(T),\\
	\epsilon(t) & = & -\frac{1}{\alpha}\textrm{Im}\big<\chi(t)|\mu|\Psi(t)\big>,\\
        \int_{0}^{T}dt\,\epsilon^{2}(t) & = & F_0.\label{controleqs}
\end{eqnarray}
which are solved iteratively~\cite{zhu,janreview} by 
applying here the forward-backward propagation scheme of 
Werschnik and Gross~\cite{Werschnik2}. Typically a converged
field is obtained within $100\ldots 300$ OCT iterations.
We have applied {\tt octopus} code~\cite{octopus} in 
all the calculations.

\section{Results}

\subsection{Superposition in a one-dimensional well}
\label{results1}

We start by considering a single particle with mass $m=1$ in a 
one-dimensional square quantum well with infinite boundaries 
and a length $L=20$. The energy eigenstates are given
by $E_n=\pi^2 n^2/(2 m L^2)=(\pi^2/800)n^2$ with $n=1,\,2,\ldots$ 
By expanding the wave function
and equalizing the phase factors it is straightforward
to show~\cite{square,square_fractional,square_revival} 
that the exact quantal revival 
time for {\em any} wave function is given by
$T_{\rm rev}=4 m L^2/\pi=1600/\pi$. However, if the wave function
is a superposition of two eigenstates with $E_n$ and $E_m$,
the first full revival occurs already at 
$T^{mn}_{\rm rev}=2\pi/(E_m-E_n)$.

In our first example our initial state is a 
superposition of the lowest two eigenstates, so that 
the first revival time
with $\left|\left<\Psi(x,T_{\rm rev})|\Psi(x,0)\right>\right|^2=1$
is $T^{12}_{\rm rev}=T_{\rm rev}/3=1600/(3 \pi)$.
Starting from the {\em same state}
but supplying our quantum well with 
a Gaussian impurity 
$V_{\rm imp}(x)=\beta\exp[-(x-x_0)^2/\gamma^2]$ with $\beta=0.1$,
$\gamma=1$, and $x_0=2.5$ leads to a reduced overlap,
$\left|\left<\Psi'(x,T_{\rm rev})|\Psi(x,0)\right>\right|^2=95.9\%$.
Here $|\Psi'\big>$ and $|\Psi\big>$ correspond to 
distorted and clean systems, respectively.

\begin{figure}
\includegraphics[width=0.99\columnwidth]{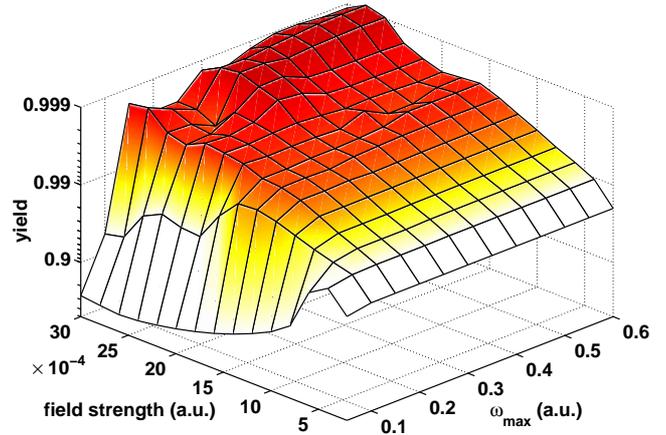}
\caption{(color online) Yield as a function of the 
field strength and maximum frequency  $\omega_{\rm max}$ of 
the optimized field in a revival process of a
one-dimensional square quantum well distorted by
a Gaussian impurity. The initial state is a superposition
of the two lowest eigenstates in a clean system.
}
\label{fig1}
\end{figure}

Our task now is to restore the full quantum revival by 
optimizing a control field with a target 
$\Phi_{\rm F}=\Psi(x,0)=\Psi(x,T)$.
Figure~\ref{fig1} shows the obtained yield (overlap) 
as a function of the maximum allowed frequency 
$\omega_{\rm max}$ and the field strength of the initial
constant field $\epsilon$ (so that the fluence 
$F_0=T_{\rm rev} \epsilon^2$ is kept fixed in the OCT procedure).
Generally, Fig.~\ref{fig1} demonstrates that the optimization
works in the desired manner,
so that we find a significant increase in the yield 
even up to $>99.9\%$. The price to pay is embedded in the
allowed intensity and frequency of the field as
analyzed in the following.

As shown in Fig.~\ref{fig1} the yield is strongly increased when
$\omega_{\rm max}$ is increased above $\sim 0.15$.
It can be expected that the "critical" frequency is
related with the energy gaps in the spectrum. Now
the critical frequency is clearly above the first few energy 
gaps, so it seems likely that (de-)excitations at relatively
high levels are needed in the distorted system
to produce a high overlap with the initial state.
The coupling with the optimized field with the 
energy levels is studied in more detail in Sec.~\ref{results2}.
We point out that at small cutoff frequencies 
($\omega_{\rm max}\lesssim 0.15$) the yield is smaller
than in the distorted system {\em without} a control field;
with a limited frequency range, and especially 
when supplied with a high field strength, the field is causing 
harm in the system despite the optimization.

In a reasonable frequency range, the required 
fluence to reach a high overlap is very small. For example,
$99\%$ yield can be obtained with 
$F_0\sim 2\times 10^{-3}$, which is only about $5\%$
of the first excitation energy $E_1-E_0=0.037$.
The energy carried by the field is also much 
smaller than the difference in the energy levels
of the system with and without the impurity.
Furthermore, in conventional control 
problems where, e.g., charge transfer or a
desired excitation is optimized,
the typical intensities are larger by several orders
of magnitude despite similar system 
characteristics~\cite{double1,double2,ring}.
Overall, we may conclude that in processes where the 
inherent dynamics gives a high initial overlap 
(as in the partial revival), OCT finds the route to 
almost $100\%$ yield with an extremely low price.

\begin{figure*}
\includegraphics[width=1.85\columnwidth]{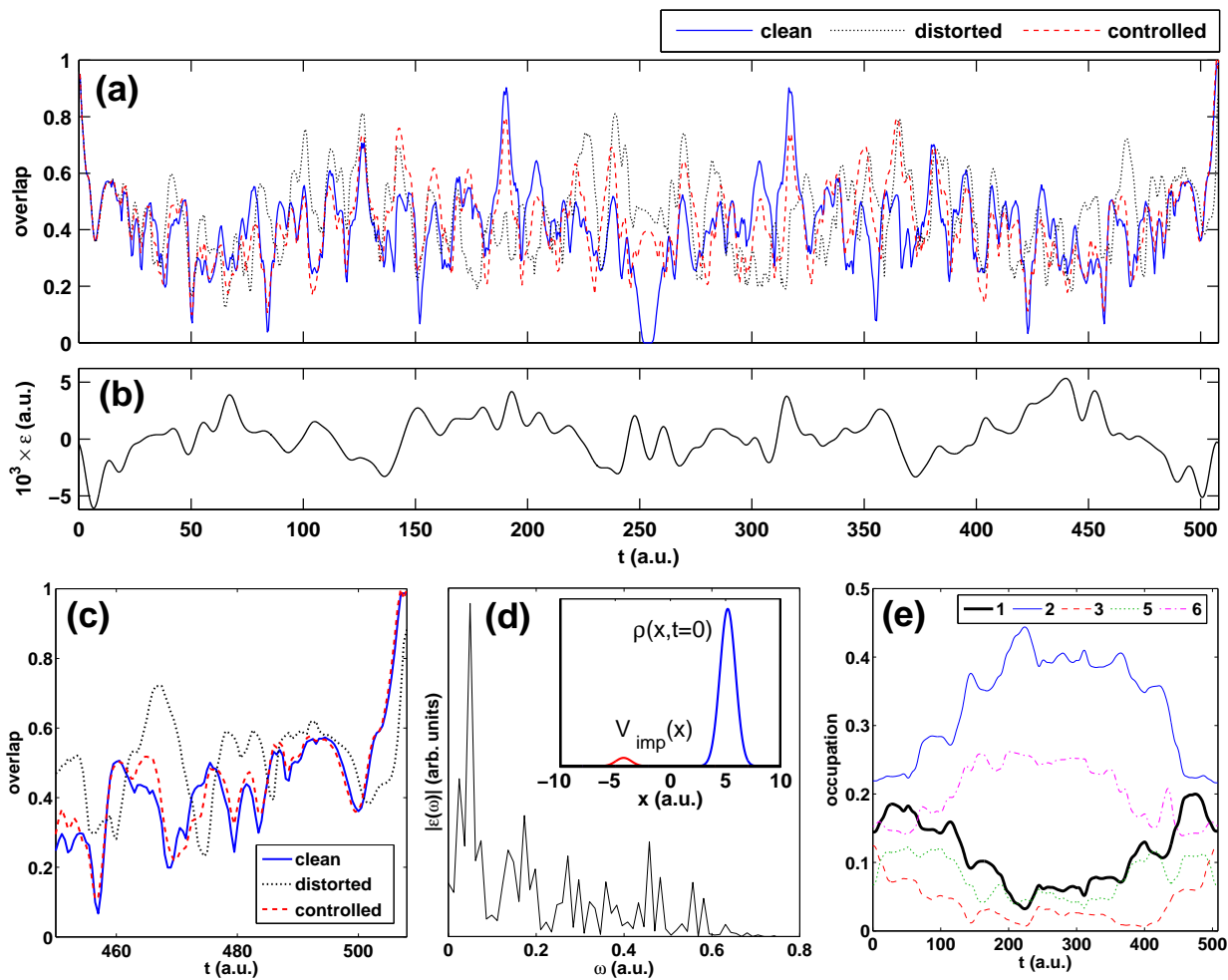}
\caption{(color online) Controlled revival of a one-dimensional 
wave packet in a square quantum well. 
(a) Time-dependent overlap of the propagated
wave packet with the initial one in a clean system (solid lines),
in a distorted system (dotted lines), and in a distorted by controlled
system (dashed lines). In (c) the last part of the propagation 
is zoomed. (b) Optimized control field. (d) Fourier spectrum of the control
field. The inset shows the external potential with the initial wave
packet and the impurity. (e) Projections of the propagated
single-electron orbitals to the initial ones.
The maximum allowed frequency is $\omega_{\rm max}=1$ and 
the fluence is fixed to $F_0=0.00203$. 
}
\label{fig2}
\end{figure*}

\subsection{Gaussian packet in a one-dimensional well}
\label{results2}

Next we study the revival of a Gaussian wave packet in
the same one-dimensional quantum well. The wave packet
is defined as $\Psi(x,0)=\exp(-(x+x_0)^2)/\sqrt{\pi}$
with $x_0=-3\sqrt{3}$, and the impurity is the same
as before [see the inset of Fig.~\ref{fig2}(d)].
As the wave packet consists of a large ensemble of
eigenstates the first revival will occur at the
universal revival time $T_{\rm rev}=1600/\pi$ as discussed in the 
previous section. This is set as the time of the 
control field, and the target is now the (single-particle) density
$\rho_{\rm F}(x,T=T_{\rm rev})=|\Psi(x,0)|^2$.

Figure~\ref{fig2} summarizes the OCT results obtained
with a threshold frequency $\omega_{\rm max}=1$ and 
fluence $F_0=0.00203$. 
Figure~\ref{fig2}(a) shows the overlap with the initial
wave packet as a function of time in the case of
(i) a clean system (solid line), (ii) a distorted system with
the impurity (dotted line), and (iii) a distorted but
controlled system (dashed line).
During the last stages of the propagation 
[Fig.~\ref{fig2}(c)]
the controlled wave packet closely follows the 
ideal system and reaches $99.9\%$ overlap, whereas
the uncontrolled evolution leads to about $88\%$. 
Thus, the optimization seems to find the correct ``path''
in time, so that the field continuously corrects the 
evolution toward the ideal one, even though the target
functional is {\em not} time-dependent. 

The optimized control field and its Fourier spectrum are shown in 
Figs.~\ref{fig2}(b) and (d), respectively. The strongest
two peaks in the spectrum at $\omega=0.025\ldots 0.05$ correspond
to the transition between the first two levels with 
$\Delta E=0.037$, and the peak at $\omega\sim 0.15$ is likely to
correspond to the transition between the 5th and 6th states with
$\Delta E=0.13$.
Figure~\ref{fig2}(e) shows the projections of the propagated
single-electron orbitals (of the system with an impurity and
a control field) to the initial orbitals (of the clean system).
Here the excitation processes between the 1st and the 2nd, and,
on the other hand, between the 5th and the 6th states are clearly
visible. At the end of the pulse the initial occupations are 
reached as the initial wave packet is reconstructed in the
controlled revival process.

\begin{figure}
\includegraphics[width=0.99\columnwidth]{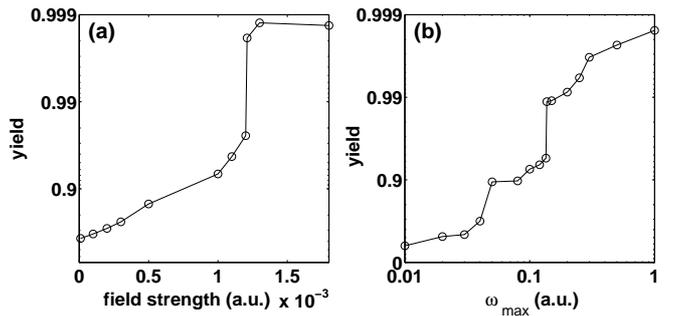}
\caption{Yield in the controlled revival of a one-dimensional 
Gaussian wave packet as a function of (a) the field strength 
(without a filter, i.e., $\omega_{\rm max}\rightarrow\infty$) 
and (b) the maximum allowed frequency
when the field strength of the initial pulse is 
$\epsilon=0.002$.
}
\label{fig3}
\end{figure}

In Fig.~\ref{fig3} we show the effect of the field strength
and the maximum allowed frequency on the obtained yield 
in the controlled revival of a Gaussian wave packet.
The critical field strength (of the initial constant field)
to reach a yield close to $0.999$ 
-- after the optimization within a fixed fluence --
is $\epsilon \sim 1.2\times 10^{-3}$ corresponding to $F_0\sim 0.00073$. 
As a function of $\omega_{\rm max}$ we find two steps at
around $0.004\ldots 0.005$ and $0.15$, respectively.
They correspond to the excitations $1\leftrightarrow 2$
and $5\leftrightarrow 6$ and confirm the importance of
those processes in the controlled evolution.

Next we consider the effect of the impurity 
characteristics on the controllability. It is expected that
a narrow impurity as the one in our example 
[inset of Fig.~\ref{fig2}(d)] probes higher excitations
than a broader impurity, and hence the dynamics and control 
procedure are qualitatively different in those cases. 
To test this, we
increase the width parameter $\gamma$ of the Gaussian impurity
from 1 to 3 in $V_{\rm imp}(x)$ (see Sec.~\ref{results1})
and decrease the height $\beta$ from 0.3 to 0.2. 
This change reduces the quality of the revival from $88\%$
to $80\%$. However, there is no significant change in the
control process: the yields are generally comparable,
and the step in Fig.~\ref{fig3}(a) is at the same place.
The yield as a function of the threshold frequency is
also similar, although the step structure is not so
prominent.

\subsection{Gaussian packet in a two-dimensional well}

\begin{figure}
\includegraphics[width=0.99\columnwidth]{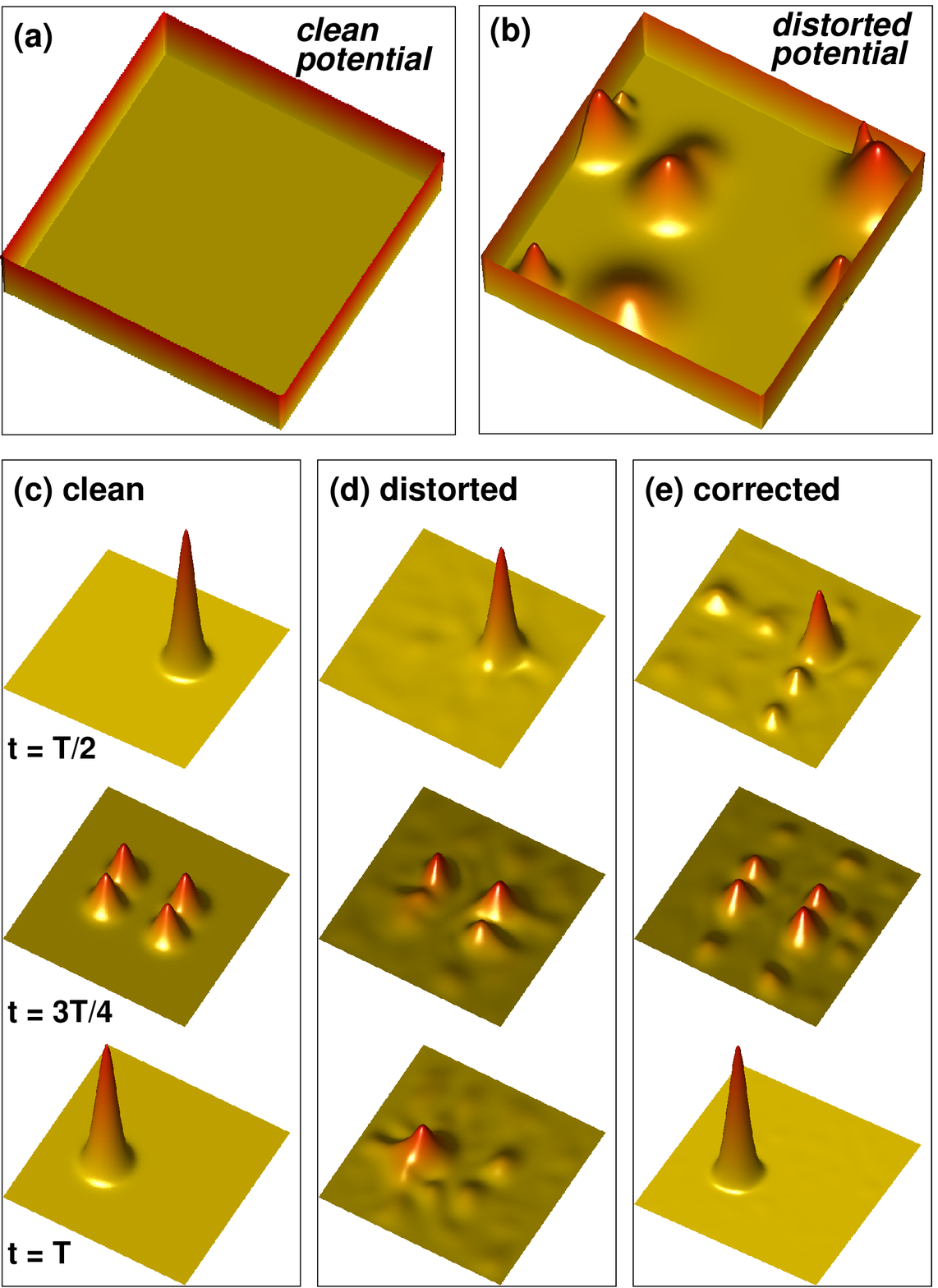}
\caption{(color online) (a-b) External potentials of clean
and distorted square quantum wells, respectively. (c-e) Single-electron
densities at times $t=T/2$, $3T/4$, and $T$ of the propagation in 
a clean, distorted, and controlled system, respectively.
}
\label{fig4}
\end{figure}

Finally we move our attention to a two-dimensional example.
This extends the physical relevance of 
the proposed scheme to two-dimensional electron gases having
a variety of applications in, e.g., quantum Hall and quantum 
dot physics~\cite{reimann}. We consider a square quantum well
with hard-wall boundaries and side length $L=12$ 
[see Fig.~\ref{fig4}(a)]. The initial Gaussian wave packet 
is given by $\Psi(x,y,0)=\delta\,\exp(-(x+x_0)^2+(y+y_0)^2)/(2 \delta^2)$
with $\delta=0.7$, $x_0=1$, and $y_0=2$.
We note that due to symmetry the wave-packet 
propagation is separable to $x$ and $y$ components. However,
we expose the system to ten {\em random} repulsive impurities visualized
in Fig.~\ref{fig4}(b) (two of them merged together), so that the 
problem is no longer separable.

\begin{figure}
\includegraphics[width=0.99\columnwidth]{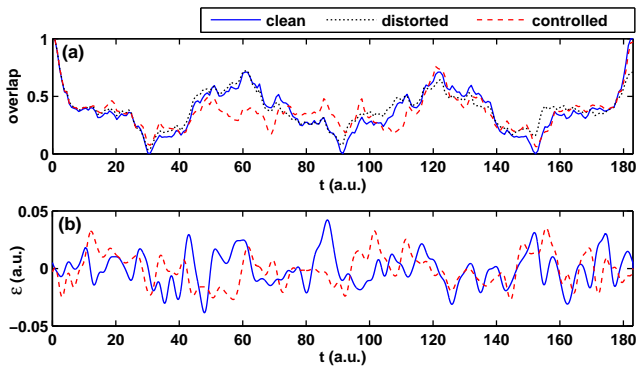}
\caption{(color online) (a) Overlap as a function of time
in a revival process of a two-dimensional Gaussian wave
packet in a square quantum well. (b) Optimized two-component
($x$ and $y$) control field leading to $97\%$ overlap.
Here $F_0=0.07348$ and $\omega_{\rm max}=0.5$. 
}
\label{fig5}
\end{figure}

Time-evolution of the wave packet in the distorted potential
leads to $71\%$ revival quality. The overlap as a function of
time is shown in Fig.~\ref{fig5}(a). To correct the process
we add a control field with two independent components polarized 
in $x$ and $y$ directions, respectively. As in the previous example,
the target is the density of the initial (single-electron) 
wave packet at $t=T=T_{\rm rev}$.
The total field fluence is fixed to $F_0=0.07348$ and the maximum 
frequency allowed in the optimization is $\omega_{\rm max}=0.5$. 
With these constraints OCT produces a control field shown in 
Fig.~\ref{fig5}(b) that drives the system to $97\%$ overlap.
Thus, the control procedure works well also in a two-dimensional
system.

Figures ~\ref{fig4}(c-e) show the single-electron densities during the
time evolution at a half of the revival time
$T/2$, at $3T/4$, and at the revival time $T$. Interestingly,
at $T/2$ the distorted packet is rather close to the ideal one
as shown as a high overlap in Fig.~\ref{fig5}(a). Towards the 
end of the evolution, however, the controlled procedure starts 
to resemble the ideal one, whereas the distorted process becomes
very different. Hence, it seems that the essential adjustments in
the controlled evolution always occur during the last stages
of the process. The relevance of the earlier stages in the
``system preparation'' could be studied by considering
{\em partial} control fields affecting
the system only during, e.g., the last $10\%$ of the time 
evolution. 
In addition, it would be worthwhile to examine if
the ideal evolution (in a clean system) can be followed
{\em continuously in time} by applying a time-dependent target 
functional~\cite{janreview}. These issues are beyond 
the scope of the present study and subject to future 
research.

Finally we comment on the simplification of using a fixed
and static impurity configurations in the above examples.
The obtained control field is found only for a
specific configuration; a general field able to control
a large ensemble of potentials would be impossible to find in
practice. Nevertheless, we point out there are several
physical situations, e.g., in semiconductor heterostructures,
where the impurity configuration in a given experimental 
setup can be found. For example, in a quantum-dot experiment in 
Ref.~\cite{jens} the size and position of a migrated impurity 
ion was accurately determined in a system that was otherwise
close to an ideal two-dimensional harmonic oscillator. In such
a situation, the Hamiltonian is known and optimization of the 
field to obtain a predefined target is possible.

\section{Summary}

To summarize, we have analyzed quantum revival processes of
single-particle states and Gaussian wave packets in one- and 
two-dimensional quantum wells. We have shown 
that the quality of the quantum revival in a realistic (distorted)
system can be greatly improved by coupling the system to an
external control field. The control field can be optimized 
with quantum optimal control theory to maximize the overlap
between the initial wave function and the time-propagated one
at the revival time -- within predefined constraints on the
fluence and maximum allowed frequency. We have analyzed how the
field constraints affect the obtained yields, and found that
very low intensities are sufficient to obtain yields above
$99\%$ in one-dimensional systems. The threshold 
frequencies to achieve high yields can be associated with excitation 
energies in the system. Our procedure has broader implications 
to general control problems, where the objective is to supplement 
the inherent dynamics disrupted by irregularities.
The demonstrated applicability to two dimensions shows that the 
proposed approach could be used to increase quantum fidelity 
in, e.g., quantum Hall devices.

\vspace{1cm}

This work was supported by the Academy of Finland,
the Wihuri Foundation, and the Magnus Ehrnrooth Foundation. 
CSC Scientific Computing Ltd.
is acknowledged for computational resources.

%
%

%

\end{document}